# Light-induced topological phase transition with tunable layer Hall effect in axion antiferromagnets

Cong Zhou, Jian Zhou[*]

*Center for Alloy Innovation and Design, State Key Laboratory for Mechanical Behavior of Materials, Xi'an Jiaotong University, Xi'an, 710049, China*

[*]Email: jianzhou@xjtu.edu.cn

## Abstract

The intricate interplay between light and matter provides effective tools for manipulating topological phenomena. Here, we theoretically propose and computationally show that circularly polarized light hold the potential to transform the axion insulating phase into quantum anomalous Hall state in $MnBi_2Te_4$ thin films, featuring tunable Chern numbers (ranging up to ±2). In particular, we reveal the spatial rearrangement of the hidden layer-resolved anomalous Hall effect under light driven Floquet-engineering. Notably, upon $Bi_2Te_3$ layer intercalation, the anomalous Hall conductance predominantly localizes in the nonmagnetic $Bi_2Te_3$ layers that hold zero Berry curvature in the intact state, suggesting significant magnetic proximity effect. Additionally, we estimate variations in the magneto-optical Kerr effect, giving a contactless method for detecting topological transitions. Our work not only presents a strategy to investigate emergent topological phases, but also sheds light on the possible applications of the layer Hall effect in topological antiferromagnetic spintronics.

Recent years have witnessed significant interests in intrinsic antiferromagnetic (AFM) materials due to their unique and switchable nature. Compared with ferromagnetic materials, AFM materials that exhibit zero net magnetizations are immune to stray field and subject to ultrafast control kinetics.[1-3] One of the most promising AFM materials with tunable van der Waals (vdW) stacking patterns is $MnBi_2Te_4$. As depicted in Figure 1a, the basic building unit is a septuple layer (SL), consisting seven atomic layers in the sequence of Te-Bi-Te-Mn-Te-Bi-Te. Within each SL, the magnetic Mn prefer ferromagnetic order with out-of-plane spin polarization, while the magnetic moments between adjacent SLs align antiparallel, resulting in a fully compensated A-type AFM ground state.[4-6] Recently, both theoretical predictions and experimental observations have revealed quantum anomalous Hall (QAH) state in odd-SL $MnBi_2Te_4$ films, and axion insulating feature with a zero Hall plateau in even-SL $MnBi_2Te_4$ films.[4, 7-10] The interlayer magnetic exchange can be weakened by intercalating $Bi_2Te_3$ quintuple layers (QLs),[11-14] as illustrated in Figure 1b. These compounds, successfully fabricated in recent years, have become ideal platforms for realizing various exotic physical phenomena due to the tunable interplay among magnetic order, topological character, and vdW structures.[5, 14-22] For example, very recent experimental efforts have proposed layer-resolved Hall effect in $MnBi_2Te_4$ films, which is a spatially distributed hidden topological feature.[8, 23, 24]

With these versatile topological properties in a single material family, it is natural to ask how they can be manipulated. Previous experiments suggest that their topology can be tuned by thickness[5, 10, 25] and magnetic configuration under external magnetic field.[11, 26, 27] Here, we predict another scheme that relies on contactless optical method to fine-tune the topological character of the fully compensated AFM system. In comparison to the previously electrical[8, 23, 28, 29] and magnetic[20, 21, 27] approaches, optical control offers good spatial resolution and ultrafast kinetics, and is more optically accessible for thin films.[28] The light effect arises from a time-dependent periodic field engineering under photon irradiation, according to the Floquet theory,[30-34] with Hamiltonian $H(t) = H(t + T)$, where $T = 2\pi/\omega$ is the time periodicity and $\omega$ the angular frequency. The solutions can be written as a product of a time-periodic function and a nonperiodic phase factor, as in the Bloch theorem for static electronic bands.[35, 36] In the Fourier expansion ansatz, the Hamiltonian and wavefunctions are $H(t) = \sum_m e^{-im\omega t} H_m$ and $|\Phi(t)\rangle = \sum_m e^{-im\omega t}|\Phi^m\rangle$, respectively, where the time-periodic function $|\Phi(t)\rangle$ is the Floquet-Bloch state. This would generate a series of replicas in frequency space, arising from the light-dressed electronic states.[36] When the laser frequency $\omega$ is much larger than interested electronic energy range, the periodic replicas become nearly energetically isolated. In such situation, the light frequency lies in the transparent frequency regime (off-resonant from direct photon absorption), so that light absorption effect can be significantly reduced. In this situation, they would renormalize the electronic Hamiltonian according to the van Vleck's expansion[37, 38]

$$H_F = H_0 + \sum_{m\neq 0} \frac{[H_{-m}, H_m]}{m\omega} + o\left(\frac{1}{\omega}\right). \qquad (1)$$

We will truncate the summation at the first order ($m = 1$), which could provide sufficient accurate results. This process can be simplified as the hybridization between the original Bloch bands ($m = 0$) and the Floquet side bands ($m = \pm 1$), forming a new Floquet-Bloch bands,[39-42] as illustrated in Figure 1c. Such Floquet-Bloch band renormalization has been observed by utilizing real-time angle-resolved photoelectron spectroscopy in various systems such as black phosphorous,[43] transition metal dichalcogenide,[44] etc.[45-50]

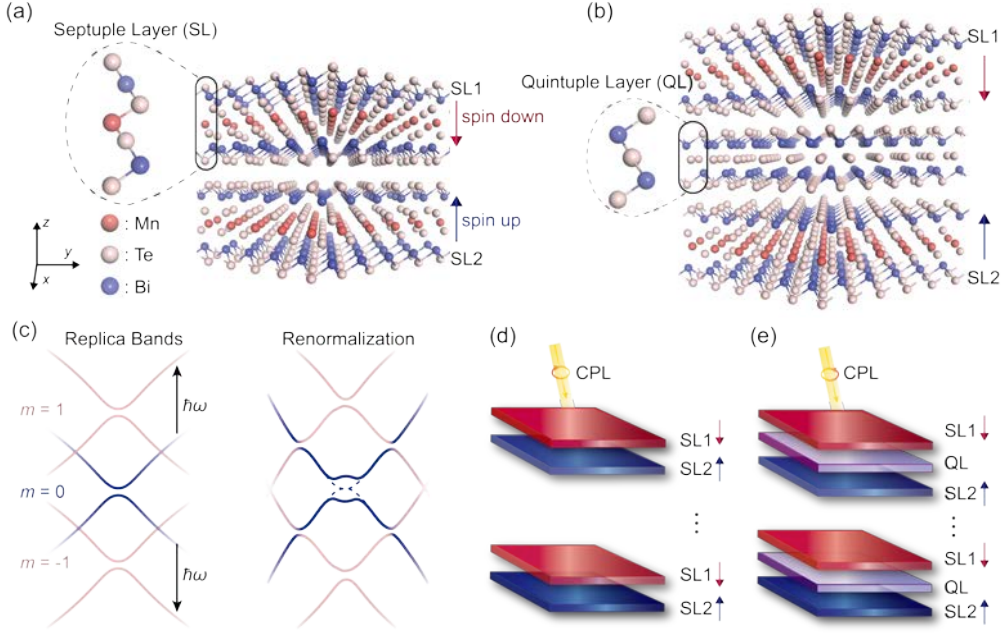

**Figure 1.** Atomic structure of stacking unit for (a) $MnBi_2Te_4$ and (b) $MnBi_2Te_4/Bi_2Te_3$. (c) Diagram illustrating band renormalization of Floquet-Bloch bands. The schematic diagram of even-SL slabs under circularly polarized light for (d) $MnBi_2Te_4$ and (e) $MnBi_2Te_4/Bi_2Te_3$.

As schematically shown in Figures 1d and 1e, we construct two types of fully compensated AFM insulator slabs. The first one consists only $MnBi_2Te_4$ bilayer (denoted as MBT). The second one includes $Bi_2Te_3$ intercalation, forming sandwich-like configuration of $MnBi_2Te_4/Bi_2Te_3$ (denoted as MBT/BT). Both these stacking patterns can be fabricated by molecular beam epitaxy scheme. Through theoretical analysis and computational exploration, we meticulously assess their light-driven electronic band dispersion variation. Moreover, we scrutinize their spatially-resolved Berry curvature distribution and layer-dependent anomalous Hall (LAH) effect. We show that under intermediate light irradiation, one can successively modulate the axion insulating phase (Chern number $\mathcal{C} = 0$) into $\mathcal{C} = \pm 1$ and $\pm 2$. We further unravel LAH distribution for each topological state. Interestingly, in the MBT/BT structure, we reveal that the BT layers dominantly contribute to the nonzero Chern index, suggesting strong magnetic proximity effect. In addition, we propose that one could measure the magneto-optical Kerr effect (MOKE), which serve as feasible physical characterization accompanied with topological transitions.

We build thick slab models of MBT and MBT/BT as illustrative examples from their bulk counterparts. The electronic structures are obtained from their bulk Hamiltonians by truncating along the ⟨001⟩ directions (see Methods). The MBT slab consists eight SLs (~10.7 nm thickness), and the MBT/BT consists eight SLs and four QLs (~15.0 nm thickness). These slab models are sufficiently thick, eliminating electronic coupling between the top and bottom SLs. We calculate their electronic band dispersion along the high symmetric $k$-path. Both the systems show direct bandgap feature, with the valence band maximum and conduction band minimum at $\bar{\Gamma}$. The bandgap values are 58.6 and 106.9 meV, respectively. Since both systems are $\mathcal{PT}$ (inversion multiplies time reversal operations) symmetric, each state is doubly degenerate from both spin up and spin down channels.[51]

To elucidate the topological transitions under normal incidence of circularly polarized light, we examine the Floquet-Bloch band evolution near the Fermi level.[52] Note that tangential incidence is ill-defined numerically, as the $z$-direction is assumed to be nonperiodic. In addition, the linearly polarized light, to the first order, does not break either $\mathcal{P}$ or $\mathcal{T}$, is unlikely to trigger Berry curvature variation. The light alternating electric field in the $x$–$y$ plane is represented by vector potential $\vec{A} = A[\eta \sin(\omega t)\hat{x} + \cos(\omega t)\hat{y}]$. Here, $A$ denotes the amplitude, and $\eta = \pm 1$ signifies the handedness of light, with +1 for right-handed circularly polarized light (RCPL) and −1 for left-handed circularly polarized light (LCPL). Thus, $\hbar\omega$ and $\frac{eA}{\hbar}$ represent the photon energy and light strength, respectively. In the following, we take a representative photon energy $\hbar\omega = 5$ eV that is much larger than the interested energy spectrum range (on the order of a few hundred meV), and it meets the off-resonant requirements for van Vleck's expansion. This value is also larger than typical interband transition, so that it would not result in significant photon absorbance (to generate electron-hole pairs) in the sample. It is noted that one cannot straightforwardly evaluate the potential nonradiative electron-hole recombination, it would be challenging to estimate the heating effect. Under laser irradiation, the stimulated radiation ratio is enhanced, which may further reduce the nonradiative process. Also, one can only apply a few laser pulses rather than continuous wave to trigger the Floquet band engineering, as in recent experimental observations.[36, 39, 40] These could suppress the potential overheating issue.

We trace the $\bar{\Gamma}$ point bandgap value of the MBT system under LCPL and RCPL, as shown in Figure 2a. One observes that light irradiation reduces the intrinsic bandgap, and the same intensity LCPL and RCPL yield the same bandgap value. The bandgap closes at $\frac{eA}{\hbar} = 0.27$ Å$^{-1}$, beyond which it reopens with finite global bandgap. At $\frac{eA}{\hbar} = 0.32$ Å$^{-1}$, the bandgap at $\bar{\Gamma}$ becomes 10.4 meV. Since CPL breaks $\mathcal{PT}$, spin splitting occurs under general $\boldsymbol{k}$ points. The spin-polarized band dispersion variation is plotted in Figures 2c–2d. It is evident that light chirality controls the spin polarization characteristics near $\bar{\Gamma}$. The bandgap close and reopen usually indicates topological phase transition. In order to explore this, we evaluate the $\boldsymbol{k}$-space Berry curvature via[53]

$$\Omega_n^z(\boldsymbol{k}) = -2\text{Im} \sum_{m \neq n} \frac{\langle \tilde{u}_{n\boldsymbol{k}} | \hat{v}_x | \tilde{u}_{m\boldsymbol{k}} \rangle \langle \tilde{u}_{m\boldsymbol{k}} | \hat{v}_y | \tilde{u}_{n\boldsymbol{k}} \rangle}{(\omega_{n\boldsymbol{k}} - \omega_{m\boldsymbol{k}})^2}, \qquad (2)$$

where $|\tilde{u}_{n\boldsymbol{k}}\rangle$ represents the Floquet-Bloch quasi-eigenstate, and its quasi-energy is $\hbar\omega_{n\boldsymbol{k}}$. $\hat{v}_i = \frac{\partial H_F}{\hbar \partial k_i}$ ($i = x, y$) is the velocity operator. Integrating the valence band Berry curvature over the first Brillouin zone (BZ) yields the total Chern number $\mathcal{C} = \int \frac{d^2\boldsymbol{k}}{(2\pi)^2} \sum_n f_n(\boldsymbol{k}) \Omega_n^z(\boldsymbol{k})$ of the slab and the anomalous Hall conductance $\sigma_{xy} = \mathcal{C}\frac{e^2}{h}$, where $f_n(\boldsymbol{k})$ is the Fermi-Dirac distribution function.[54, 55] The phase diagram of Chern number as a function of the CPL handedness and intensity is depicted in Figure 2a. Before light is illuminated, the total Chern number is zero, consistent with the axion insulating nature. When the light intensity exceeds a critical value ($\frac{eA}{\hbar} \approx 0.27$ Å$^{-1}$), LCPL (RCPL) would induce QAH state with Chern number of +1 (–1). The momentum space distribution of Berry curvature is plotted in Figures 2e–2f. One observes that pronounced peaks (positive or negative, depending on light handedness) emerge in the vicinity of $\bar{\Gamma}$, consistent with the band inversion feature.

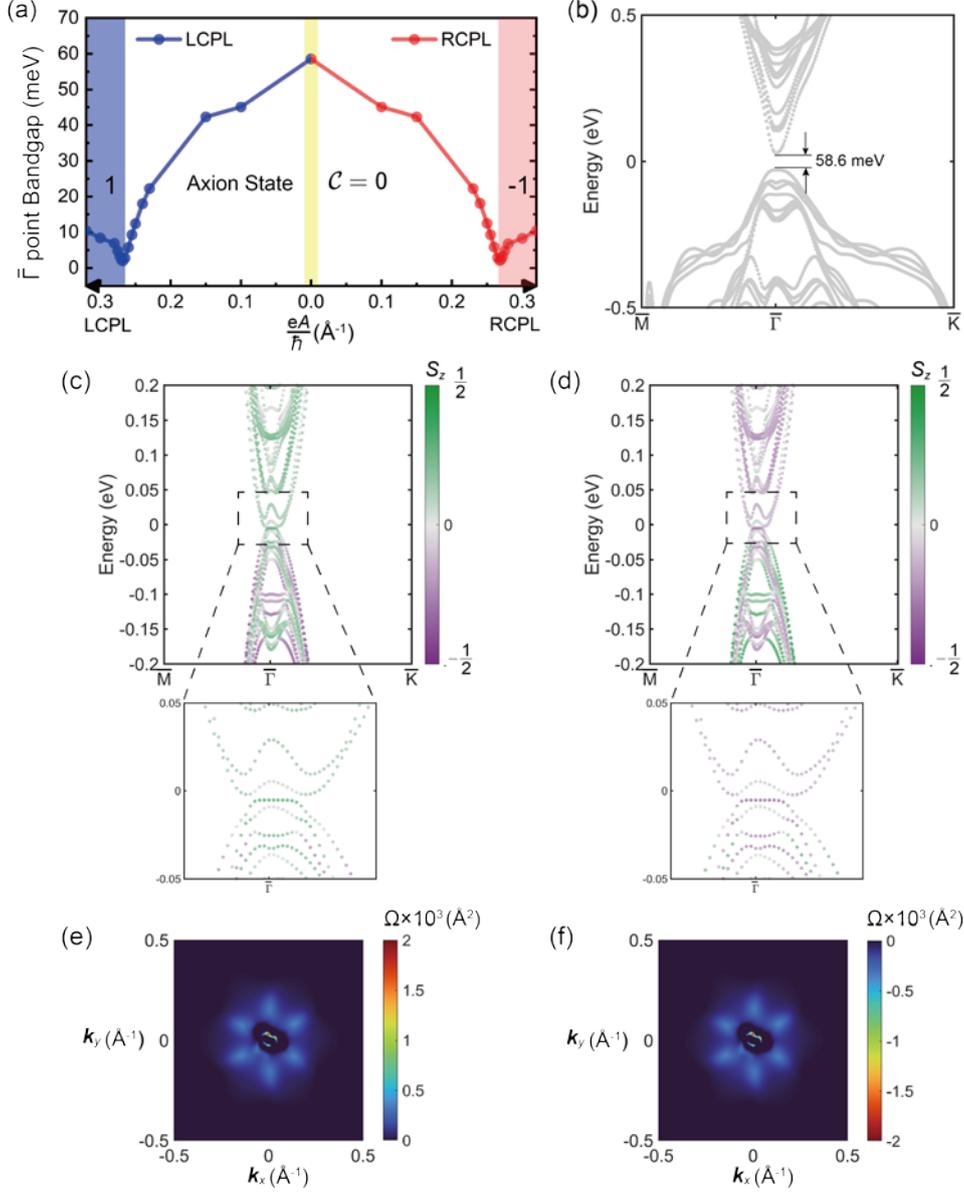

**Figure 2.** Electronic behavior in MBT system. (a) Bandgap at $\bar{\Gamma}$ as a function of the LCPL and RCPL intensity, along with the light-controlled phase diagram of MBT. The axion insulator region is represented in yellow. The QAH phase regions, symbolized by $\mathcal{C} = +1$ and $\mathcal{C} = -1$, are depicted in light-blue and light-red shades, respectively. The electronic band dispersion (b) before light is irradiated, (c) under LCPL and (d) under RCPL. The insets are enlarged plots near $\bar{\Gamma}$. Colorbar represents $\langle S_z \rangle$ for each state. The momentum space distributions of Berry curvature under (e) LCPL and (f) RCPL. In (c)-(f), the light intensity $\frac{eA}{\hbar} = 0.32$ Å$^{-1}$. The band dispersion at transition point is plotted in Figure S1.

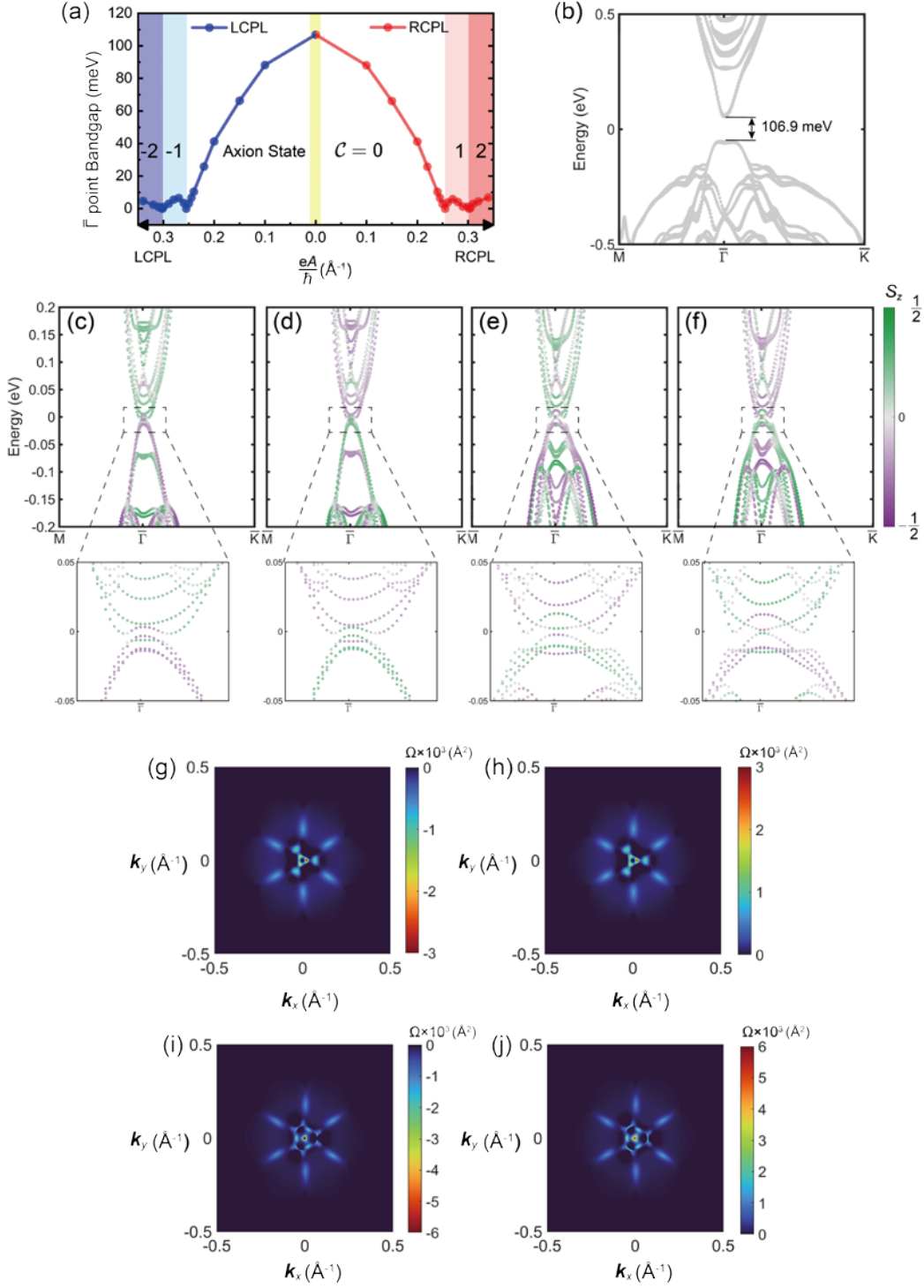

**Figure 3.** Electronic behavior for MBT/BT. (a) Bandgap at $\bar{\Gamma}$ as a function of the light intensity, along with the light-controlled phase diagram of MBT/BT. Yellow area signifies axion insulating state, while different QAH phase regions with Chern numbers of $\mathcal{C} = -1, +1, -2, +2$ are indicated. The electronic band dispersion under different CPL conditions, (b) without light, (c) and (e) under LCPL, (d) and (f) under RCPL. The $k$-dependent Berry curvature, (g) and (i) under LCPL, (h) and (j) under RCPL. In (c)-(d) and (g)-(h), light intensity $\frac{eA}{\hbar} = 0.27$ Å$^{-1}$. In (e)-(f) and (i)-(j), light

intensity $\frac{eA}{\hbar} = 0.32$ Å$^{-1}$. The band dispersions at transition points are plotted in Figure S2.

As for MBT/BT, our calculations reveal similar photon-dressed bandgap reduction, as shown in Figure 3a. Interestingly, here we find two gap close-and-reopen processes, at $\frac{eA}{\hbar} = 0.25$ Å$^{-1}$ and $0.30$ Å$^{-1}$, respectively. Chern number calculations reveal that they correspond to $\mathcal{C} = \pm 1$ and $\mathcal{C} = \pm 2$, suggesting high Chern number states driven by intermediate light. In the $\mathcal{C} = \pm 1$ regime, the largest bandgap is 6.5 meV at $\frac{eA}{\hbar} = 0.27$ Å$^{-1}$, and the bandgap could reach 6.9 meV when $\mathcal{C} = \pm 2$ ($\frac{eA}{\hbar} = 0.34$ Å$^{-1}$). These values are comparable with the calculated results for odd layered MnBi$_2$Te$_4$ slabs, hence the topologically nontrivial QAH effect can be experimentally observed. Here we do not require large magnetic field or thickness engineering to realize QAH state. We plot the typical band dispersion in Figures 3c-3f, showing twice of band inversion at the Fermi level. The Berry curvature dominantly appears near $\bar{\Gamma}$, as plotted in Figures 3g–3j. Contrary to the MBT situation, in the MBT/BT slabs, LCPL (RCPL) yields negative (positive) Chern numbers. This indicates that the LAH effect may differ in the two systems.

In order to explore the spatial distribution of anomalous Hall conductance, we calculate the hidden LAH effect. The LAH effect has been proposed and discovered in multilayer MnBi$_2$Te$_4$ thin films, aiding to characterize their axion insulating phase. Theoretically, it is associated with layer-resolved local Berry curvature distribution. In order to evaluate it, we introduce a spatial projection operator $P_l = \sum_{i \in l} |\psi_i\rangle\langle\psi_i|$.[56] Here, $|\psi_i\rangle$ are localized orbitals (e.g., in the Wannier representation), and the summation runs over all orbitals centered on the $l$-th layer. Then the layer-dependent Berry curvature can be evaluated by inserting $P_l$ in the Kubo perturbation formulism. We plot the layer-resolved anomalous Hall conductance (LAHC) $\sigma_{xy}^l$ ($l = 1 - 8$) for 8-SL MBT slab (Figure 4a). Before light is applied, $\sigma_{xy}^l$ shows a clear odd-even oscillation feature. Note that the LAH does not localize only on the topmost and bottom layers, and finite $\sigma_{xy}^l$ enters the inner of the slab. Previous works[57] have suggested that when the thickness of the slab exceeds five SLs, the LAH conductance on surfaces is between $0.42 \frac{e^2}{h}$ and $0.35 \frac{e^2}{h}$, no longer exactly half-quantized. Our results provide consistent results, giving $\sigma_{xy}^{l=1} = -\sigma_{xy}^{l=8} = 0.42 \frac{e^2}{h}$. Note that here the top (bottom) layer possesses spin down (up) polarization. If time-reversal $\mathcal{T}$ is applied, the distribution of LAH would also flip. The interior layers contribute relatively weaker LAH conductance (about $\pm 0.23 \frac{e^2}{h}$). The $\mathcal{PT}$ symmetry assigns opposite LAHC distribution, namely, $\sigma_{xy}^l = -\sigma_{xy}^{9-l}$. These results agree very well with its axion insulating feature.

Under LCPL dressed $\mathcal{C} = +1$ state, the $\sigma_l$ distribution changes drastically. As plotted in Figure 4a, the LAH still shows weak odd-even oscillation, but with an overall incremental trend. In detail, compared with the intact state, there exhibit significant variations of LAH conductance in the top (from 0.42 to $-0.19 \frac{e^2}{h}$) and bottom layers (from $-0.42$ to $0.40 \frac{e^2}{h}$). The Berry curvature aligning with LAH conductance mainly localizes at the bottom surface (Figure 4c). Opposite trends can be seen under RCPL. This is because different handedness is $\mathcal{PT}$ symmetric, so that one has $\sigma_{xy}^l(\text{LCPL}) = -\sigma_{xy}^{9-l}(\text{RCPL})$. Therefore, we suggest that different chiral light could effectively regulate the LAH effect in MBT thin films.

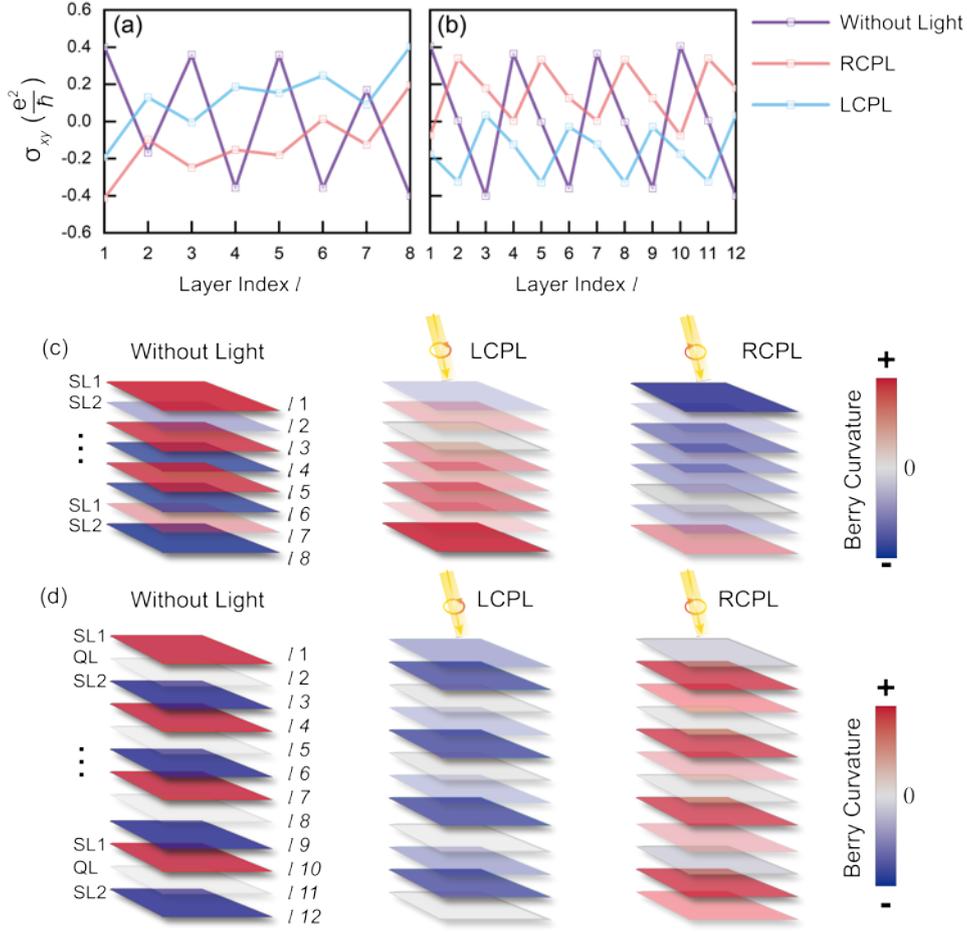

**Figure 4.** Layer-resolved characterizations of various topological states in an 8-SL MBT and MBT/BT slabs. Layer-resolved anomalous Hall conductance as a function of layer index $l$ in, (a) MBT, (b) MBT/BT. The violet, light-blue and light-pink lines represent in $\mathcal{C}=0,+1,-1$ state, respectively. Schematic illustration of the local Berry curvature distribution under various conditions in (c) MBT and (d) MBT/BT. Colorbar shows local Berry curvature variation. Light intensity for the QAH state is $\frac{eA}{\hbar} = 0.32 \text{ Å}^{-1}$.

The intercalation of $Bi_2Te_3$ QLs drastically alter the LAH distribution. In the absence of laser illumination ($\mathcal{C} = 0$), the spin up and spin down $MnBi_2Te_4$ SLs exhibit opposite LAH signs, while the nonmagnetic $Bi_2Te_3$ QLs show almost zero values (Figure 4b). In detail, the top and bottom SLs give $\pm 0.4 \frac{e^2}{h}$. Such LAH distribution considerably changes under CPL illumination. We plot the total $\mathcal{C} = -2$ state ($\frac{eA}{\hbar} = 0.32 \text{ Å}^{-1}$) in Figure 4d, leaving the similar $\mathcal{C} = -1$ case in Figure S3. As depicted in Figure 4b, the dominant LAH conductance distribution now lies in the Bi$_2$Te$_3$ layer, each giving $-0.33\frac{e^2}{h}$. Consequently, the accumulated LAH conductance of all Bi$_2$Te$_3$ QLs decreases from zero to $-1.3\frac{e^2}{h}$ (Figure S4). This indicates that the four Bi$_2$Te$_3$ layers carry 67% anomalous Hall conductance, while the eight MnBi$_2$Te$_4$ layers give the rest 33%. Furthermore, all spin-down SLs (SL1) contribute a total of $-0.6\frac{e^2}{h}$, while the LAC conductivity contributed by spin-up SLs (SL2) is nearly zero. When the chirality of light is flipped, these contributions are also reversed.

The light-dressed band variation and topological phase transition could be detected via their (layer-resolved) Hall conductance variations. Here, in addition to the electric signal, we propose that one can also adopt noncontacting optical measurements using a weak light. Since the incident pump light breaks $\mathcal{PT}$, the magneto-optical Kerr effect (MOKE) would serve as a sensitive technique to probe the topological phase transition.[58, 59] The real and imaginary parts of the complex Kerr angle refer to Kerr optical rotation angle and the Kerr ellipticity of the reflected polarization, respectively, $\phi_K = \theta_K + i\epsilon_K$. In polar geometry, the complex Kerr angle of a sample with higher than threefold rotational symmetry is[60]

$$\theta_K(\omega) + i\epsilon_K(\omega) = \frac{-\sigma_{xy}(\omega)}{\sigma_{xx}(\omega)\sqrt{1 + \frac{i\sigma_{xx}}{\varepsilon_0 \omega}}}. \tag{4}$$

Here, $\varepsilon_0$ is dielectric constant in vacuum. The frequency-dependent dielectric function $\varepsilon_{ij}$ and optical conductance $\sigma_{ij}$ are based on Kubo perturbation theory[61, 62]

$$\varepsilon_{ij}(\omega) = 1 - \frac{e^2}{\varepsilon_0}\int \frac{d^2\mathbf{k}}{(2\pi)^2} \sum_{m\neq n} f_{nm} \frac{\langle \tilde{u}_{n\mathbf{k}}|\hat{v}_i|\tilde{u}_{m\mathbf{k}}\rangle \langle \tilde{u}_{m\mathbf{k}}|\hat{v}_j|\tilde{u}_{n\mathbf{k}}\rangle}{E_{m\mathbf{k}} - E_{n\mathbf{k}} - \hbar\omega - i\eta} \tag{5}$$

$$\sigma_{ij}(\omega) = \frac{e^2}{\hbar}\int \frac{d^2\mathbf{k}}{(2\pi)^2} \sum_{m\neq n} f_{nm} \frac{\langle \tilde{u}_{n\mathbf{k}}|\hat{v}_i|\tilde{u}_{m\mathbf{k}}\rangle \langle \tilde{u}_{m\mathbf{k}}|\hat{v}_j|\tilde{u}_{n\mathbf{k}}\rangle}{(E_{m\mathbf{k}} - E_{n\mathbf{k}})^2 - \hbar(\omega + i\eta)^2}. \tag{6}$$

Here, $f_{nm} = f_n - f_m$ refers to occupation difference between band $n$ and $m$ at momentum $\mathbf{k}$. $\hbar\eta$ is the broadening factor which phenomenologically describes impurity, electron-phonon interaction, disorder, and environmental effects. In principle, it depends on both $n$ and $k$, but a thorough evaluation is not straightforward and computationally challenging. Hence, we follow conventional approaches to take a universal value ($\hbar\eta = 0.1$ eV). Our results for both MBT and MBT/BT films are plotted in Figure 5.

At the intact axion insulating state, both $\epsilon_K$ and $\theta_K$ vanish under $\mathcal{PT}$ symmetry. Under Floquet theory driven QAH state, finite MOKE signals emerge. One sees that the magnitude of both $\theta_K$ and $\epsilon_K$ generally increase as stronger pump CPL is applied. The LCPL and RCPL with the same intensity generate opposite MOKE angles. For MBT, the LCPL yields negative $\theta_K$ and positive $\epsilon_K$. For example, at the probe light photon energy of 1.5 eV (wavelength of ~800 nm), $\theta_K$ (and $\epsilon_K$) could reach –0.88 (and 0.47) mrad in MBT ($\mathcal{C} = 1$ at $\frac{eA}{\hbar} = 0.32$ Å$^{-1}$), which can be experimentally detected. For the MBT/BT, increasing Chern number could generally enhance the magnitude of $\theta_K$ (and $\epsilon_K$). At 1.5 eV incident photon energy, $\theta_K$ (and $\epsilon_K$) becomes 0.19 (and 0.14) and 0.38 (and 0.20) when $\mathcal{C}$ is tuned to be 1 and 2, respectively. This arises from the fact that the $\mathcal{C}$ corresponds to the static limit of optical Hall conductance $\sigma_{xy}(\omega)$, which scales MOKE angles according to Eq. (4).

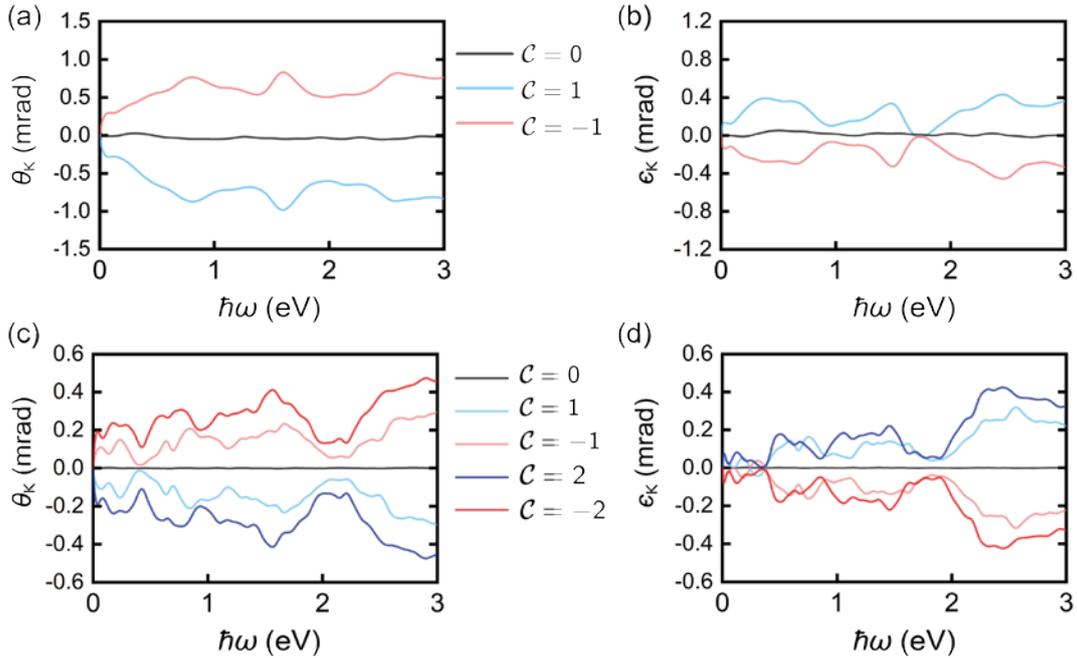

**Figure 5.** The calculated complex Kerr angles for the MBT (upper panels) and MBT/BT (lower panels): Kerr angle $\theta_K$ (left panels) and Kerr ellipticity $\epsilon_K$ (right panels). The $\mathcal{C}=0$ state correspond to intact $\mathcal{PT}$ system. Generally, under lower CPL pump would give smaller Kerr angles.

Very recently, Xu *et al.* used 500–1000 nm CPL to trigger AFM Néel vector switch in MnBi$_2$Te$_4$ between two degenerate AFM states.[28] It stems from the effective axion field of light ($\boldsymbol{E} \times \boldsymbol{B}$), and the phase transition is based on thermodynamic imbalance between the two AFM configurations. The magnetoelectric coupling breaks both $\mathcal{P}$ and $\mathcal{T}$, which is essential for the observation. In this work, the photon energy is much higher, and the AFM switch is unlikely to occur. Hence, we can focus on the topological phase transition and the evolution of LAH effect.

We would like to point out that due to some theoretical and computational uncertainties, the predicted field strength for the topological transition may quantitively deviate from realistic experimental measurements. One of the main reasons is the well-known bandgap underestimation. If the incident photon energy varies, the critical light intensity may also alter. On the other hand, as the Berry curvature is a robust quantity, the (layer-resolved) anomalous Hall conductance remains the same.

Note that Eq. (1) is only valid when the incident photon energy is larger than the interested energy regime. Hence, here we take a representative value (5 eV). If we slightly reduce (or enhance) the photon energy, the critical field for topological phase transition would decrease (or increase), but the main physical picture remains. One may wonder if such a high photon energy will emit electron. Actually, considering that the spot size of laser is usually a few tens micrometers (containing ~$10^{12}$ unit cells), and one can estimate that a few femtosecond laser pulses would emit ~$10^{-3}$ – $10^{-2}$ electrons per unit cell. This is marginal for rigid band approximation and can be compensated by substrate effect or gate voltage. Hence, the main physical conclusion would not change.

The topological phase transition in MnBi$_2$Te$_4$ family is usually achieved via adjusting the thickness and interlayer magnetic configuration (under magnetic field), and the LAH has been detected under gate voltage. Our work presents another contactless approach for such a phase transition, and our LAH calculation suggest that the magnetic proximity effect from MnBi$_2$Te$_4$ SLs to Bi$_2$Te$_3$ QL is significant. These await further experimental verifications and quantum manipulation applications.

In this work, utilizing first-principles calculations in combination with Floquet theory, we systematically investigate the light-induced topological transitions and MOKE signal in MBT and MBT/BT slabs. In the MBT films, the CPL field can engineer the original axion insulating state $\mathcal{C} = 0$ into a QAH state $\mathcal{C} = \pm 1$, depending on light handedness. For the MBT/BT, high Chern number topological states can be achieved. More strikingly, we reveal that the LAH effect dominates in the BT QLs that is otherwise silent without light, suggesting significant magnetic proximity effect. In addition to the real-time angular resolved photoelectron spectroscopy to detect the band engineering, the (layer resolved) Hall conductance offers an electrical characterization approach. This work offers new perspectives on the universal implementation and control of the QAH and LAH effects, guiding the potential applications of spatial texture of Berry curvature in topological antiferromagnetic spintronics.

**Methods**

We perform first-principles density functional theory calculations in the Vienna *ab initio* simulation package (VASP)[63, 64] and the generalized gradient approximation (GGA) with the Perdew-Burke-Ernzerhof (PBE) type exchange-correlation potential is adopted.[65] The projector augmented wave (PAW) method[66] is used to treat the core electrons, and the valence electrons are expanded using a plane wave basis set with a kinetic cutoff energy of 350 eV. The Hubbard $U$ correction is adopted to treat the strong correlation in the magnetic $d$ orbitals and the effective Hubbard $U$ parameter is chosen to be 5.34 eV for the Mn-$d$ orbital.[9, 51, 67] The vdW interactions are empirically included using the Grimme's D3 scheme.[68] Spin-orbit coupling (SOC) is always included self-consistently, if not stated otherwise. The Monkhorst-Pack special ***k***- meshes[69] for MBT and MBT/BT are set to be (14×14×2) and (14×14×1), respectively. The total energy and force component convergence criteria are set to be $1 \times 10^{-7}$ eV and $1 \times 10^{-5}$ eV/Å, respectively. In order to simulate the electronic structures of slab systems, we use Wannier functions[70-73] to fit the tight-binding model (based on the Mn-$d$, Bi-$p$, and Te-$p$ orbitals) from the counterpart bulk materials. Then we truncate along ⟨001⟩ directions to build the Hamiltonian of slab models. The Wannier function basis set (a total number of $J$) can be written as

$$|n\bm{R}\rangle = \frac{1}{N_k}\sum_{\bm{k}} e^{-i\bm{k}\cdot\bm{R}} \sum_{m=1}^{J} U_{mn}|m\bm{k}\rangle \tag{7}$$

Here, $U_{mn}$ is a unitary transformation for Bloch state $|m\bm{k}\rangle$, $N_k$ is the total number of the $\bm{k}$-mesh, and $\bm{R}$ is the vector measured from the home cell $\bm{R}=\bm{0}$. The interaction between different function is $\widetilde{H}_{mn}^{\bm{R}} = \langle m\bm{0}|H|n\bm{R}\rangle$. To construct the Hamiltonian in thin films (with $N_\perp$ unit cells in the normal $z$-direction), we perform Fourier transformation in the $x$ and $y$ directions ($\bm{k}_\parallel$) and truncate the interaction above and below the slab. The Hamiltonian is a $(N_\perp J \times N_\perp J)$ matrix, and its component is

$$H_{(i_\perp,m),(j_\perp,n)}(\bm{k}_\parallel) = \sum_{\bm{R}} \widetilde{H}_{mn}^{(\bm{R}_\parallel, R_\perp)} e^{i\bm{k}_\parallel \cdot (\bm{R}_\parallel + \bm{\tau}_n - \bm{\tau}_m)} \tag{8}$$

Here, we denote $\bm{R} = (\bm{R}_\parallel, R_\perp)$, and the layer index number $j_\perp = i_\perp + R_\perp$, with $1 \leq i_\perp, j_\perp \leq N_\perp$ and $1 \leq m, n \leq J$. $\bm{\tau}_{n,m}$ is the Wannier wavefunction center position. Since the MBT and BT layers are vdW stacked, one does not need to include the dangling bond and surface construction effects. With the constructed Hamiltonian, a dense $\bm{k}$-mesh grid of (600×600×1) is adopted to integrate the first BZ to obtain the optical conductivities, which has been tested to achieve sufficient converged accuracy.

**Acknowledgments.** This work is supported by National Natural Science Foundation of China under grant number 12374065. The Hefei Advanced Computing Center is also acknowledged where the computations are performed.